# Demonstration of Latency-Aware 5G Network Slicing on Optical Metro Networks

B. Shariati[1,*], L. Velasco[2], J.-J. Pedreno-Manresa[3], A. Dochhan[3], R. Casellas[4], A. Muqaddas[5], O. González de Dios[6], L. Canto[6], B. Lent[7], J.E. López de Vergara[8], S. López-Buedo[8], F. Moreno[9], P. Pavón[9], M. Ruiz[2], S. Patri[3], A. Giorgetti[10,11], F. Cugini[10], A. Sgambelluri[10], R. Nejabati[5], D. Simeonidou[5], R.-P, Braun[12], A. Autenrieth[3], J.-P. Elbers[3], J. K. Fischer[1], R. Freund[1]

*1 Fraunhofer HHI, Berlin, Germany*
*2 Universitat Politècnica de Catalunya, Barcelona, Spain*
*3 ADVA, Munich, Germany*
*4 Centre Tecnològic Telecomunicacions Catalunya, Castelldefels, Spain*
*5 University of Bristol, Bristol, UK*
*6 Telefónica, Madrid, Spain*
*7 Qognify GmbH, Bruchsal, Germany*
*8 Naudit HPCN, Madrid, Spain*
*9 Universidad Politécnica de Cartagena, Cartagena, Spain*
*10 CNIT, Pisa, Italy*
*11 IEIIT-CNR, Pisa, Italy*
*12 Deutsche Telekom, Germany*

*Corresponding author: behnam.shariati@hhi.fraunhofer.de*

The H2020 METRO-HAUL European project has architected a latency-aware, cost-effective, agile, and programmable optical metro network. This includes the design of semi-disaggregated metro nodes with compute and storage capabilities, which interface effectively with both 5G access and multi-Tbit/s elastic optical networks in the core. In this paper, we report the automated deployment of 5G services, in particular, a public safety video surveillance use case employing low-latency object detection and tracking using on-camera and on-the-edge analytics. The demonstration features flexible deployment of network slice instances, implemented in terms of ETSI NFV Network Services. We summarize the key findings in a detailed analysis of end-to-end quality of service, service setup time, and soft-failure detection time. The results show that the round-trip-time over an 80 km link is under 800 μs and the service deployment time under 180 seconds. © 2021 Optical Society of America

http://dx.doi.org/10.1364/JOCN.99.099999

## 1. INTRODUCTION

5G promises to deliver low-latency high-capacity end-to-end (e2e) services in a very short time upon the customer's request [1]. Meeting such a goal entails not only extending the optical network to the metro to provide low-latency high-capacity services [2], but also from a sophisticated solution able to dynamically instantiate e2e network slices with minimum service setup time. Once a set of connectivity and computing resources need to be set up and assigned to each network slice. Once in operation, such a network should be monitored to detect degradations and anticipate failures, leaving enough time to implement adequate countermeasures (e.g., tuning device parameters or re-route connections) that avoid any disruption (see [3-7]).

Video surveillance for public safety is a time-critical and high-bandwidth vertical use case of growing concern in the society. In fact, internet video surveillance traffic has increased significantly in the past, reaching 3.4% of all internet video traffic in 2021 [8].

In a very simplified view, video surveillance applications are made up of IP cameras that are connected to recording servers. One server might have 100 to 250 cameras connected, either fixed cameras, thermal cameras to enable surveillance by night and Pan-Tilt-Zoom (PTZ) cameras to follow objects. Client applications request live or archive video from multiple servers simultaneously. Video analytics are performed on camera or on the recording server; analytics can range from simple motion detection to face and even behavior recognition. All these functions of video surveillance require various amounts of bandwidth, storage, and computational power. One of the key requirements to perform intelligent video surveillance is the availability of powerful computation resources in the surveillance zone to perform video analytics. One solution that does not require deploying expensive computational resources in the video cameras consists in outsourcing computational tasks to edge computing nodes in the network operator's infrastructure. Video analytics running in the edge can elaborate on the video footages and then the results should become available in the surveillance zone within a fraction of second. Such solution requires a high-capacity connectivity to allow streaming the high-definition video footages from the surveillance zone towards the edge computing nodes, which might be distributed across a city. Moreover, the whole lifecycle of such solution should not impose a significant latency as it may compromise the accuracy of the system in terms of response time. In addition, low latency is a key factor in operating a tracking Pan-Tilt-Zoom (PTZ) camera remotely.

In this context, the H2020 METRO-HAUL project [9] has designed and built a *smart* optical metro infrastructure able to deliver requirements for time-critical and high-bandwidth vertical use cases, including public safety [10]. The developed solution comprises several hardware and software pieces that together form a unique networking infrastructure, including both packet and optical layers, along with an integrated Network Function Virtualization (NFV) and disaggregated network orchestration, which provides advanced connectivity and ETSI Network Services (NS) encompassing computing, storage, and networking resources [11].

This paper extends [12] and presents the control, orchestration, and management (COM) architecture that supports several key features, including e2e latency-awareness, as well as monitoring and data analytics capabilities, to operate a partially disaggregated edge computing enabled metro optical network. The main objective is to demonstrate and validate Key Performance Indicators (KPI) for delivering low-latency high-capacity services. Specifically, we demonstrate that (1) optical metro networks need to provide flexible low-latency high-capacity connections to enable real-time video surveillance and analytics; (2) network slicing allows the allocation of adequate computational resources to run the video management and analytics on remote edge servers; and (3) distribution of functionalities on edge computing nodes can be performed in a latency-aware networking scenario. Notably: (*i*) some of the video management modules are assigned to run very close to the video surveillance zone as they do not require high computational power; (*ii*) the analytics module can be placed in a more powerful computation network slice in the nearest available datacenter; (*iii*) the distribution of the video management and analytics can further enhance the network resource utilization as it does not require all the video footages to be transported to the analytics module; and (*iv*) the latency in the network slices (particularly the Round-Trip-Time -RTT) can be monitored by employing dedicated active probes. The KPIs to be demonstrated focus on the different phases of the setup (Optical connection service setup time (KPI-1), e2e connection service setup time (KPI-2), and e2e network service setup time (KPI-3), and on the time to detect degradations (KPI-4).

The demonstration includes the full workflow of planning, orchestration, deployment, and running a network slice including computing resources, as well as connectivity over an edge-computing enabled metro optical networking setup at Fraunhofer HHI in Berlin.

The remainder of the paper is organized as follows. Section 2 introduces the METRO-HAUL infrastructure, the designed COM system, and the specifics of the video surveillance application. Section 3 describes the demonstration, starting from the deployed architecture, the solution implemented for RTT measurements, and the workflow. Section 4 presents the experimental setup and the obtained results for the defined KPIs. Finally, Section 5 draws the main conclusions.

## 2. METRO-HAUL INFRASTRUCTURE, CONTROL PLANE, AND VIDEO APPLICATION

### A. Metro Infrastructure

The METRO-HAUL infrastructure spans nodes residing in Central Offices in different geographic locations, where every node combines networking, processing, and storage resources. Such modular nodes are composed of different components operating at different layers and technologies, and of different vendors realizing hardware and software disaggregation. METRO-HAUL nodes implement layer 0-1 (optical domain) and layer 2 transmission and switching (frame domain) and include Edge Computing capabilities provided by a local pool of computers to instantiate Virtualized Network Functions (VNF) with configurable amounts of processing, memory, and storage. Two specializations of the generic METRO-HAUL nodes are: a) Access Metro Edge nodes (AMEN) to interface with heterogeneous access technologies (5G and optical); and b) Metro Core Edge nodes (MCEN) nodes as gateways towards the core transport network and comprise core-oriented capabilities (Fig. 1). The nodes are controlled by a Node Agent based on NETCONF/YANG handling the integration of such disaggregated components.

### B. Control, Orchestration, and Management

The previously described metro infrastructure requires a complex COM system, which includes several subsystems and interfaces among them. The COM system augments the concept of network control plane with standard interfaces operating across domains to ensure vendor inter-operability. The architecture of the METRO-HAUL COM system has evolved from its initial design and successive refinements have been made driven by feedback gathered after implementation and integration activities. The main components include the following (see Fig. 1).

1) The *NFV Orchestrator* (NFVO) that performs Service Orchestration (involving the functional split of the service into/amongst different VNFs and their logical interconnection) and Resource Orchestration dealing with the allocation of resources to support the VNFs and the logical links. In the context of METRO-HAUL, a network slice consists of a NS deployed using the NFVO spanning multiple nodes and network domains; the VNF placement functionalities are provided by the Network Planner. The Virtual Infrastructure Manager (VIM) is the responsible for the management of the NFV Infrastructure (NFVI) and the instantiation of the Virtual Machines (VM) of the VNFs in a single datacenter domain (AMEN/MCEN).

2) The *WAN Infrastructure Manager* (WIM) is used by the NFVO to orchestrate network resources and it is responsible for the provisioning of connectivity paths between VNFs. The WIM architecture is hierarchical and aligned with IETF ACTN [13], with an SDN control per technology domain. Running on top of the SDN hierarchy, the parent SDN controller abstracts the underlying complexity and presents virtualized networks to their customers. At the optical transport layer, two approaches have been considered based on two main disaggregation models: a) Partial Disaggregation: Open Line System (OLS) and Multi-Vendor Transponders (TP) (the one considered for this demonstration), and b) Full Disaggregation: Wavelength Division Multiplexing (WDM) transport system. Regarding the control of the Passive Optical Networks (PON), it is realized through the adoption of an abstraction scheme which represents the PON as an NETCONF-enabled switch. The parent's and packet SDN controllers' northbound interfaces (NBI) are based on RESTCONF [14], using the YANG Layer 2 Service Model (L2SM) defined in [15].

On the other hand, ONF TAPI [16] was chosen to interface the NBI of the SDN controller and OLS. TAPI enables the provisioning of constrained connectivity services and the operation of an OLS. Recent versions of TAPI (2.1.2 and later) have largely increased support for the photonic media layer, allowing a new degree of flexibility in the provisioning of digital (ODU/OTU) and optical (OTSi/media channels) connectivity services. RESTCONF is the underlying protocol between the optical SDN controller and the OLS controller. The implementation shall support a set of common aspects such as: *i)* the ability to retrieve the set of Service Interface Points (SIPs) along the tunability constraints associated to each client port; *ii)* the ability to retrieve the topology of the network in terms of links and nodes, node edge points, and *iii)* the creation and deletion of optical channel connectivity services

In addition, the OpenConfig drivers included in ONOS have been re-engineered to dynamically create and configure the required logical channels starting from a blank transponder configuration and in accordance with the OTN hierarchy.

3) The *Monitoring and Data Analytics* (MDA) [17], responsible for implementing autonomic networking. The monitoring system has the capability to do measurements on the data plane and to generate data records that are collected and analyzed by the MDA subsystem to discover patterns (knowledge) from the data. In METRO-HAUL, the MDA is distributed [18] and consists of MDA agents that run in the network nodes and are responsible for monitoring data collection, aggregation, and knowledge usage. Aggregated monitoring data is conveyed to the MDA controller. From collection, data can feed Machine Learning algorithms [19], which can be used to issue re-configuration/re-optimization recommendations towards COM modules, such as an SDN controller or orchestrator (see, e.g., [20-21]).

4) The Placement, Planning, and Reconfiguration Subsystem (*Network Planner*), responsible for optimizing the resource allocation to effectively provision services featured by heterogeneous requirements and for applying different policies and strategies. This task comprises the provisioning of VNFs in specific METRO-HAUL nodes, and the allocation of network resources [22].

## C. Video Surveillance

A smart city video surveillance application requires multiple IP cameras distributed across the whole city and connected to several recording servers. In a city-wide installation, the controlling hardware and software will be located in a central datacenter. In a larger urban area, this central data center will not be in the same network node as the servers, they will be distributed across a city and connected by the optical metro network. Fig. 2 shows the structure of the network with two network nodes and cameras. The servers act as Core System Slave (CSS) for the video management system and as Device Manager (DM) to control the cameras. They are connected to a Core System Master (CSM) server, for management and video analytics.

A client control center requests video footages from the individual servers and controls cameras manually. Live or archive video could be requested from multiple servers simultaneously, requiring high bandwidth through the optical network. For automated and manual control of the cameras, triggered by the analytics or the user, low RTT (few ms) is mandatory. Analytics can range from simple motion detection to face and even behavior recognition. All these functions of video surveillance require various amounts of bandwidth, storage, and computational power.

The following scenario is common: one or more thermal cameras are mounted on the top of a building overlooking the surrounding parking lot or entrance area. Closer to ground level, PTZ cameras track a person detected by the camera located on top of the building. The thermal camera detects the person and tracks her/his movement. The PTZ camera helps identify the person and the individual activities; the maximum RTT is 10-50 ms. In a small installation all the controlling and recording software will be located in the building that is being monitored.

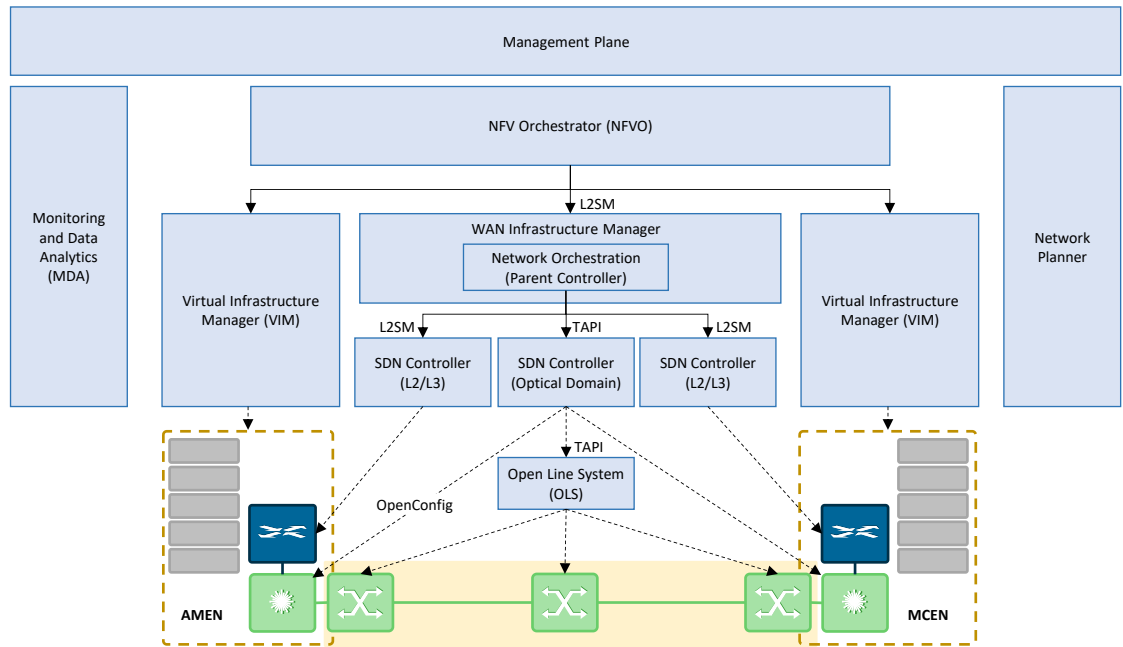

**Fig. 1.** METRO-HAUL Control, Orchestration, and Management system and network, compute, and storage infrastructure

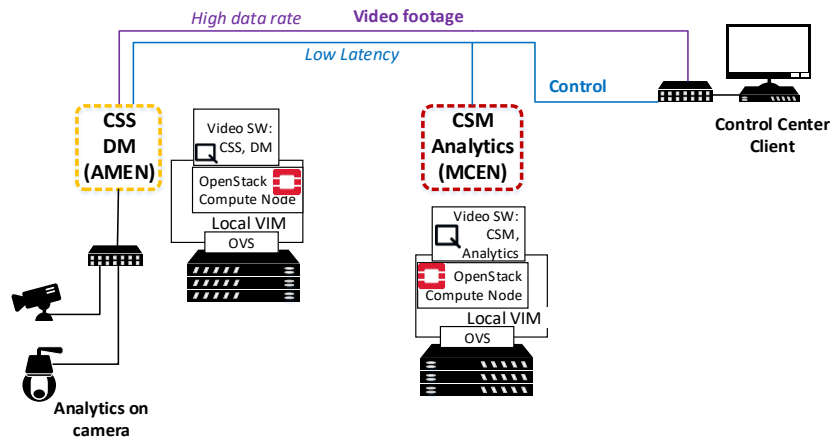

**Fig. 2.** Video surveillance scenario. Multiple cameras are connected an AMEN node. Analytics and management are performed at MCEN.

The other factor that a metro network contributes to city surveillance is bandwidth. Particularly, applications including face detection or license plate recognition demands high bandwidth, as the recognition software requires high resolution transmitted via MJPG rather than H-264 or H-265. This is even more important if the objective is to detect faces in crowds, since typically 100 pixels resolution is necessary between the eyes of a face to be recognized from a large list of faces. For Full HD the bandwidth consumption is 76 Mbit/s per channel, for 4K resolution it is 240 Mbit/s per channel. For a large city with 2000 channels using face recognition, 2000 channels using license plate recognition and 150.000 video channels, the total bandwidth requirement is over 700 Gbit/s. To show a fluid display of the video in real time the packet delay variance needs to be below the frame rate. Frame rates in practice vary from very few frames/s for a parking lot to 30-50 frames/s inside a casino. In a network of this size the recording and video analysis could either reside in a central data center or be distributed in computing centers in the network nodes themselves.

With the above in mind, two different use-cases are considered in this demonstration: *i)* autonomous object tracking using the PTZ camera; and *ii)* zone-crossing detection. These use-cases are described next:

**Autonomous Object Tracking**: As for this use-case, the analytics software runs on the thermal camera and a notification is sent to the PTZ camera to follow the object in the surveillance zone. The workflow of the use-case is as follows: 1) One person enters the surveillance zone; 2) the analytics that runs on the thermal camera detects the movement and provide a trajectory to the PTZ; 3) PTZ cameras follow the object (i.e., the person in our case) as long as it remains in the surveillance zone; 4) the client can watch the video footage from a remote place (i.e., the Client PC in this demonstration); and 5) the client could (potentially) interrupt steering and starts manual control of PTZ camera using the joystick, which also requires low latency connectivity to be done properly.

Zone Crossing Detection: As for this use-case, the analytics runs on one of the VNFs hosted on the remote compute nodes. The video footages recorded from the fixed cameras should be transferred to the analytics software, which is pre-configured to identify the zones. The workflow of the use-case is as follows: 1) One person walks around in the surveillance zone; 2) the video footage are streamed towards the analytics software that runs on the VM; 3) once the person crosses the identified zone border on the analytics software, an alarm is triggered; 4) once the alarm is triggered, the client can steer the PTZ manually (it can be configured that the PTZ camera is refocused to that spot).

## 3. DEMONSTRATION

### A. Deployed Architecture

Fig. 3 shows the demonstration architecture. We setup a partially disaggregated optical network comprising commercial equipment from ADVA [23]: three 2-degree semi-filterless reconfigurable optical add-drop multiplexers (ROADM), based on wavelength blockers and splitters, connected in a ring topology, and two coherent transponders (QuadFlex™). Each one of the transponders is connected, via an aggregation switch, to a compute node that acts as an edge datacenter.

The COM system relies on recent advances of the SDN and NFV paradigms. Specifically: *i)* Open-Source MANO (OSM) [24] is used as an NFVO. OSM is an open-source MANO system which is based on ETSI NFV Information models; *ii)* Open Network Operating System (ONOS) [25] is used as SDN controller; *iii)* OpenStack [26] is used as the VIM. In addition, Net2Plan [27] has been used as network planner.

Based on those existing frameworks, extensive adaptation work to the specific requirements of the project has been carried out. This includes the applicability to Metropolitan networks, the deployment of disaggregated optical networks, the importance of monitoring, telemetry and data analytics, and the interest in externalizing the algorithmic aspects (network optimization, function placement, resource allocation) to dedicated subsystems. In this work, we follow a partially disaggregated scenario, which requires a hierarchy of optical SDN controllers. Specific drivers have been developed in the optical SDN controller to manage the ROADMs through the OLS controller, as well as for transponders.

Regarding the MDA subsystem, an MDA controller [17] carries out fault degradation analysis and RTT measurement retrieval upon a request from the parent controller.

Our infrastructure also includes three cameras and their corresponding distributed Video Management System (VMS), which together carry out real-time video surveillance.

### B. Active Measurements

Another key novelty is the Quality of Service (QoS) measurement capability, which becomes possible by an active 100G probe [28] that provides real-time network measurements. Active measurements are not advised during network operation, because they increase the load of the network. However, they are useful for commissioning testing [29], when such measurements can verify whether the provided QoS meets the requirements in terms of RTT, jitter, throughput, and packet loss. Assuming high quality NSs being provided, it is of paramount importance to measure them, with high precision. To that end, our deployed probe provides RTT measurements with ns precision, throughput up to 100 Gbit/s with kb/s precision, and packet loss rates with $10^{-6}$ precision.

In order to achieve these precise measurements, we have implemented the active probe using a Field Programmable Gate Array (FPGA). In particular, the ADM-PCIE-9V3 High-Performance Network Accelerator card, which includes two 100GE interfaces, 8 GB of DDR4 memory and a Virtex Ultrascale+ XCVU3P-2-FFVC1517 FPGA. An integrated 100G Ethernet interface is in charge of communicating the FPGA side with the physical network. The FPGA internally works with a bus of 512-bit width of data and is clocked at 322 MHz (3.1 ns) to reach the needed throughput, even with the smallest frames. A packet train technique [30] has been implemented in the FPGA at 100 Gb/s, where packets are timestamped at the transmitter and at the receiver; receiver timestamps are useful to calculate the achieved throughput, whereas transmitter timestamps allow precise calculation of RTT with the clock implemented in the FPGA, without synchronizing transmitter and receiver.

The FPGA development was split into two independent designs, which reside in the same FPGA. In one hand, the transmitting side, a synthetic packet generator has been developed. On the other hand, the receiving side is in charge of filtering the packets, analyzing them and generating a summary. Both elements are briefly described next.

The Synthetic Packet Generator has been written in Verilog and implemented as a finite state machine, so its behavior is completely deterministic, and it provides the most accurate measurements. This module generates UDP datagrams with useful information for the measurement. The datagrams can be generated at the maximum throughput, up to 100 Gb/s in 100G Ethernet links. When a measurement is requested, some of the fields in the Ethernet frames can be set, such as the VLAN id, source and destination IP addresses, source and destination ports, packet size, and Bit Error Rate Test type used for the payload. Trains can be up to $2^{32}-1$, so packet loss in even very reliable links could be identified. Note, however, that this long measurement could take several minutes.

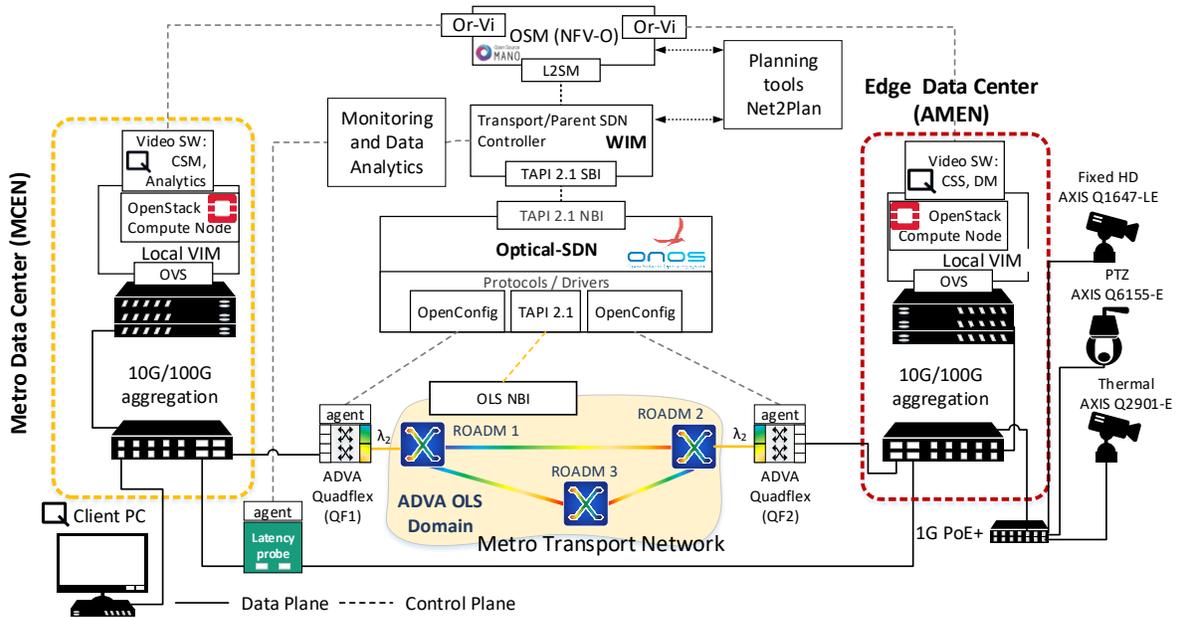

**Fig. 3.** Architecture of the Demonstration

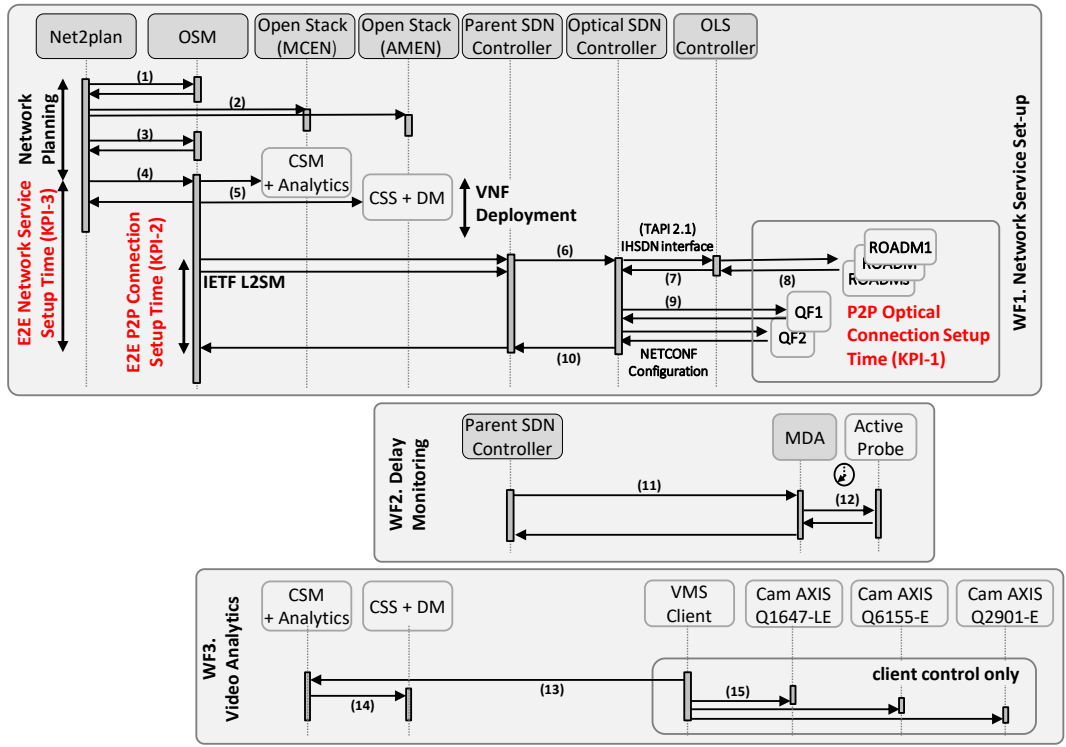

**Fig. 4.** Workflows of the Demonstration

At the receiver side, UDP datagrams are timestamped when they reach the FPGA side. The packet handler module filters packets by protocol, e.g., ARP and UDP are considered in this implementation. After that, UDP datagrams are parsed, and configurable fields are verified. The useful information is extracted to compute the packet train parameters in the statistics generator. The results of the measurements are then returned to the server once a packet train has been completely received or after a timeout.

### C. Workflow

Fig. 4 shows the interactions implemented for the demonstration, which include three main workflows executed sequentially: WF1) *Network Service Set-up*: this workflow includes iterations among different subsystems in the COM, as well as between the COM and the infrastructure in the data plane, e.g., the ROADMs. KPIs 1-3 are related to this workflow; WF2) *RTT monitoring*: it includes iterations among the active probe, the MDA controller, and the parent SDN controller; and WF3) *Video Analytics*: this workflow includes iterations among the cameras in the surveillance zone and the corresponding VNFs instantiated by OpenStack. One of the VNFs hosts the CSM and the video analytics of the VMS, while the other one hosts the CSS and the DM.

WF1 starts when the instantiation of a NS is received. The request includes a user-configurable NS descriptor stating the VNFs. Additionally, latency requisites can be defined via the Net2Plan UI. Then, the planner computes the optimal VNFs placement. The resource allocation algorithm implemented uses *i*) VNF requisites, in terms of CPU, memory, and storage, retrieved from the NS descriptors in OSM (message 1 in Fig. 4), and *ii*) status information retrieved from the OpenStack instances in the MCEN and AMEN nodes (2). Those VIMs with not enough idle resources to accommodate at least one of the VNFs, are discarded. Then, the algorithm computes the ranking of k-minimum cost service chains (SCs) on a graph where links are weighted with the RTT of the shortest path between the link end nodes in the optical transport network and the vertices with VIMs are tagged with those VNFs that they can instantiate [27]. The VNFs are placed according to the first feasible SC in the rank. Note that a ranked SC may be unfeasible if more than one VNF need to be placed in the same VIM. If there is no valid SC, or the minimum-RTT SC is above the service requirements, the request is blocked. If a valid SC is obtained, the planner sends a NS instantiation request to OSM (3, 4) with the placement information. This triggers the NFVO to request the instantiation of the VNFs. In the demo, the VNFs run different pieces of the video management system including the video analytics engine and data manager, as well as a connectivity slice to the Parent Controller (5).

The VNFs are deployed by the OpenStack instances running in the MCEN and AMEN nodes. In parallel, the parent controller requests optical network connectivity to the optical SDN controller, which proceeds to requests the connectivity from the OLS controller in charge of the ROADMs, and the two optical transponders with an OpenConfig interface (6-10).

The interaction between ONOS and the OLS consists of a number of steps. First, ONOS retrieves the TAPI Context, which includes the detailed topology of the OLS. Given the flexibility of the TAPI model, it is the choice of the OLS controller to export the actual topology in terms of ROADMs or to abstract the OLS as a single forwarding domain node. Next, ONOS retrieves the active connections or media channels that are currently provisioned in the OLS. Then, ONOS requests the creation of a media channel between the involved SIPs corresponding to the selected OLS client ports, and includes other needed data, such as the optical bandwidth and the selected frequency slot.

Transponder configuration also consists of a number of steps involving several NETCONF messages between the controller and the transponder: *i*) creation of the OTU4 logical channel is initiated by ONOS, which triggers creating the OCH and the ODU4 logical channels and mapping the ODU4 channel within the OTU4 channel and that in the OCH channel; *ii*) ONOS configures the optical frequency and the transmission power of the OCH channel; *iii*) ONOS creates a transceiver component to be associated with the client port; *iv*) ONOS creates the logical channel on the client side, which includes creating the ODU4 channel on the client side configuring the assignments from the transceiver up to the ODU4 channel; and *v*) ONOS configures the assignment between the ODU4 channels on the client and the line sides.

Once the NS has been setup, WF2 is executed to perform commissioning testing; in this case, the focus is on RTT monitoring. The parent controller starts WF2 by requesting to measure the performance of the just-created packet circuits (11); the circuit IDs, VLAN IDs, maximum RTT, and other details are included in the request. The MDA controller requests the measurements to the active probes that use the VLAN IDs for tagging the generated Ethernet frames (12). To measure the performance of a circuit, the active probe in the source of the circuit injects trains of Ethernet frames that are received by the active probe in the destination, which loops back them to the sender. Once the performance of the circuits is measured, the obtained values are replied to the MDA controller, which compares to the defined maximum and confirms the parent controller. Performance values are stored in the MDA controller, so the operator that can verify the performance of the set-up circuits.

Once the NS enters in operation, WF3 runs, and the video management system is initiated, and the VMS client requests permission to the CSM to access the cameras (13), video streams are requested to the DM (14), and cameras are configured (15).

## 4. EXPERIMENTAL SETUP AND RESULTS

### A. Experimental Setup

Fig. 5 shows the data plane for the demonstration setup at the Fraunhofer HHI premises in Berlin (Germany). The data plane includes one AMEN and one MCEN nodes, interconnected by a 3-node metro network formed by three ADVA micro-ROADMs. The cameras are connected to the AMEN node consisting of one Mellanox switch and an edge computer with an OpenStack deployment. The live traffic of the cameras and the VNF (instantiated on the compute node) is transferred through the Mellanox switch towards QuadFlex™. The client port of the transponder is connected to the switch via a 100 GbE QSFP28 transceiver and an optical cable. At the line side, the QuadFlex™ transponder sends a one-wavelength 100 Gbit/s DP-QPSK modulated signal, which then gets transported over the metro network and is received by Quadflex1 on the MCEN. The internal architecture of both AMEN and MCEN nodes is identical, where MCEN instantiates the other VNF and provides connectivity to the client PC hosting the client of the video management system. Moreover, the active probe is also connected to the Mellanox switches to measure the RTT.

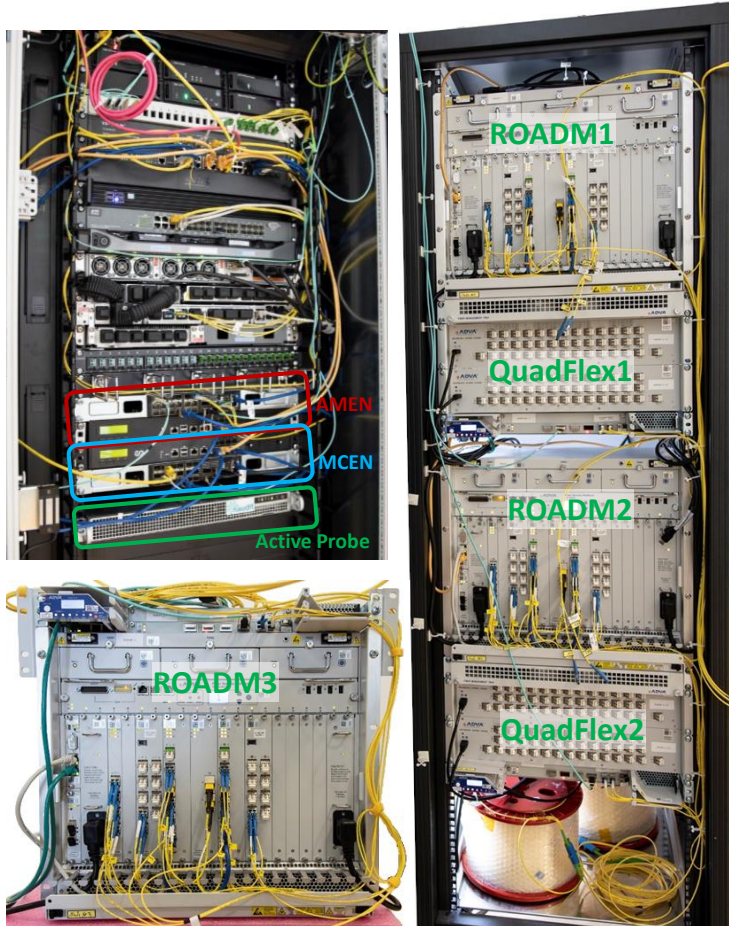

**Fig. 5.** Data Plane Demonstration Setup

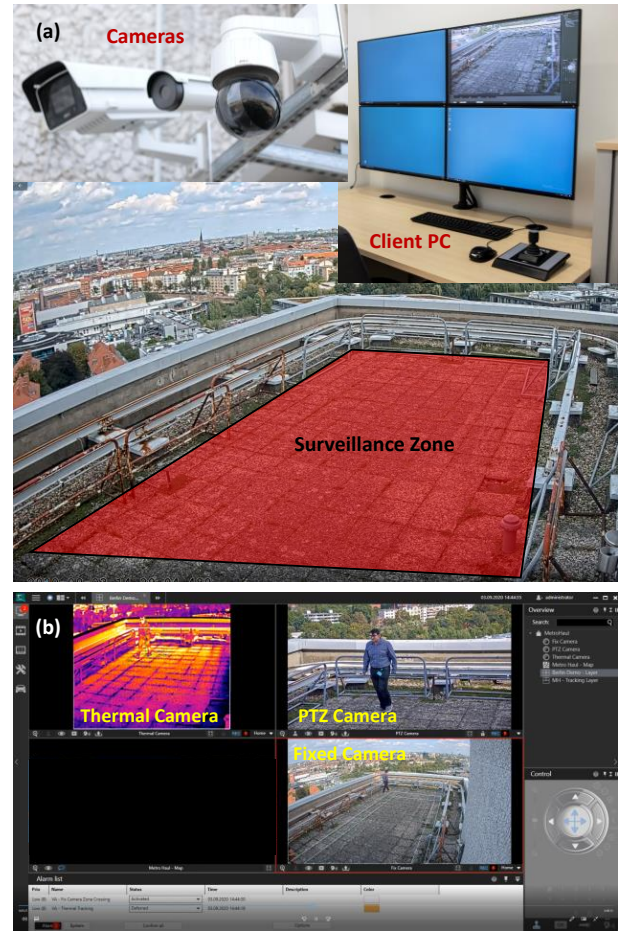

**Fig. 6.** Public Safety Surveillance Demonstration Zone

**Table 1.** QoS Measurements

| Path length [km] | Two-way Propag. [µs] | RTT [µs] | Delta [µs] | Throughput (Mb/s) | Packet loss rate |
|---|---|---|---|---|---|
| 2E-3 | 0.02 | 0.86 | 0.84 | 97,154.7 | 0 |
| 1.5E-3 | 0.015 | 2.1 | 2.1 | 97,155.0 | 1.0E-6 |
| 2.1E-3 | 0.021 | 15.3 | 15.2 | 97,153.7 | 2.1E-6 |
| 41.4 | 405.04 | 420.4 | 15.4 | 97,145.1 | 1.4E-6 |
| 80.0 | 783.16 | 799.1 | 16.0 | 97,149.0 | 1.8E-6 |

The demonstration is performed using different fiber spools between ROADM1 and ROADM2 and their impact on the performance of the set-up are measured; in particular, on the data plane performance metrics and the incurred RTT. The fiber spools are: 41 and 80 km and 2.1 m for both directions.

The rooftop of HHI was used for the set-up of the surveillance demonstration zone, as shown in Fig. 6a. Fig. 6b shows the user interface of the Qognify client that streams the video footages coming from the three cameras, i.e., the thermal camera, the PTZ cameras, and the fixed camera.

## B. Results

To assess the performance of the system, we report some of the key results obtained during the demonstration. From the data plane perspective, the optical transponders multiplex the client traffic into a 100 Gb/s DP-QPSK signal. The signal is then transmitted from ROADM1 to ROADM2. The link length between ROADM1 and ROADM2 was varied for measuring RTT. Next, we present the evaluated service setup time, QoS metrics, and soft-failure degradation time.

**Service Setup Time**: we report the experimental results for the three KPIs that characterize the service deployment time defined in Section 1. We repeated the experiments 28 times to obtain a statistically reliable value for each KPI. The graphical definition of each KPI with a detail of the different phases and the measured values are provided in the diagram in Fig. 7 and summarized in the inset. Note that, while the configuration of the optical transponders (QFs) is performed in ~2 s, their transmitting laser takes a significant time (~125 s) to warm up and stabilize. Excluding the transponders, the e2e service setup time takes ~50 s.

**QoS:** Measurements from different set-ups are summarized in Table 1 for different fiber lengths. The first three set-ups were configured for calibration purposes. The first set-up connects the probe in loopback, to estimate the latency added by the probe itself. The second one connects the aggregation switches with a direct link to estimate the introduced latency. The third set-up uses the optical equipment connected though a 2m fiber to measure the latency added by the optical path. The fourth and fifth set-ups use longer optical paths with about 41 and 80 Km, respectively.

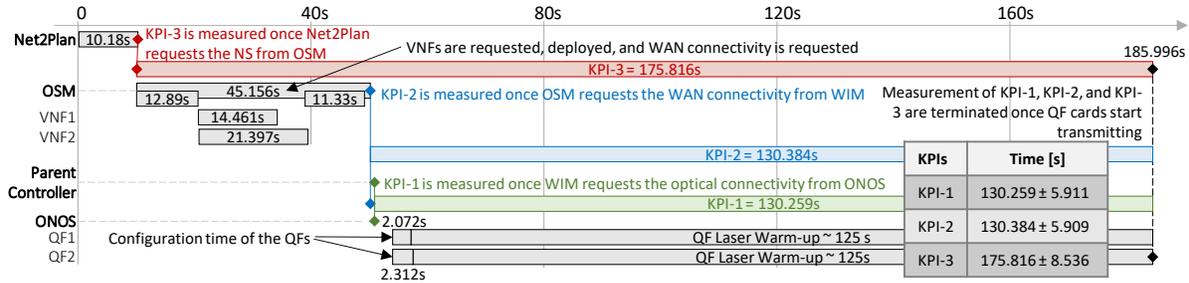

**Fig. 7.** Service set-up times (KPI-1, KPI-2, and KPI-3) of different stages of the demonstration.

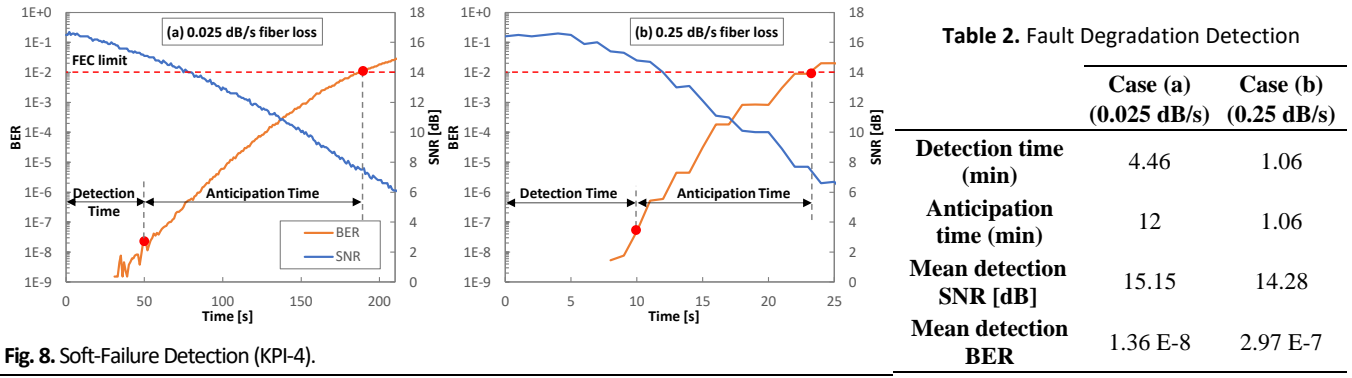

**Fig. 8.** Soft-Failure Detection (KPI-4).

**Table 2.** Fault Degradation Detection

| | Case (a) (0.025 dB/s) | Case (b) (0.25 dB/s) |
|---|---|---|
| **Detection time (min)** | 4.46 | 1.06 |
| **Anticipation time (min)** | 12 | 1.06 |
| **Mean detection SNR [dB]** | 15.15 | 14.28 |
| **Mean detection BER** | 1.36 E-8 | 2.97 E-7 |

Every single set-up was tested ten times. In each measurement, a packet train with $10^6$ packets was sent. We observe that the network throughput is near to 100 Gbit/s in all cases, with small fluctuations. Note that we computed throughput at the IP network layer, i.e., excluding Ethernet headers, preamble and interframe gap. The measurement is remarkably close to the theoretical maximum data rate that could be achieved (97.196 Gbit/s). In addition, very few packets were loss (if any) in every test. We verified in the first set-up that the probe does not lose any packets; the aggregation switches seem to be responsible for some of the loss, as tested in the second case, whereas the rest could be caused by the optical network (remaining set-ups). Nonetheless, communication was very reliable; above 99.999% in all cases. Regarding latency, results are different depending on the scenario: column "Delta" in Table 1 contains the difference between the measured RTT and the two-way propagation delay. We derive the latency budget of the different elements in the set-up by considering the three first measurements in Table 1. Specifically, *i*) 0.84 μs are added by the probe and its optical transceivers; *ii*) 1.29 μs are added by the aggregation switches and their transceivers; and *iii*) 13.1 μs are added by the METRO-HAUL optical devices. With respect to jitter (not shown in the table), it was 4 ns in the first setup, and about 7-8 ns in the rest of scenarios.

**Soft-failure Detection Time**: Here, the intention is to anticipate QoT degradation, leaving enough time to implement countermeasures (see [4-6]). We introduced intentional fiber loss by an additional variable optical attenuator to emulate the failure, firsty by 0.025 dB per second (a) and then by 0.25dB per second (b). The effects were measured in terms of SNR and pre-FEC BER at the receiver.

Illustrative examples of measured degradations and detection and anticipation times are presented in Fig. 8 for cases (a) and (b). In addition, Table 2 presents the average value for all the measurements (6 for each case). Although the detection time depends on the speed of the degradation, the anticipation time leaves at least more than one minute to reconfigure the degraded optical connection, even under rapid degradations.

All these measurements together demonstrate that the infrastructure meets the QoS goals defined for the METRO-HAUL project, proving that the developed infrastructure is able to support highly dynamic scenarios providing high bandwidth, high reliability, and low latency network services, meeting the most stringent QoS requirements of 5G networks.

## 5. CONCLUDING REMARKS

The experimental demonstration of a video surveillance vertical use-case in a latency-aware metro network with e2e service setup and network slicing including connectivity and virtual network functions has been reported in this paper. A multi-partner, multi-component integration based on software and hardware elements, including Net2plan, OSM, ONOS, and OpenConfig-enabled open terminals was achieved. The demonstration showed the integration of an NFV MANO architecture with a hierarchical, multilayer SDN control plane for disaggregated networks, including a dedicated OLS controller.

The experimental setup allowed the evaluation of the system performance, including the dynamic service provisioning, the latency-awareness, and the subsequent monitoring and fault detection processes. Four KPIs have been experimentally quantified highlighting the key benefits and asserting the key role of the optical technology for services with stringent latency and bandwidth requirements.

**Funding.** This work has been partially supported by the European Commission (METRO-HAUL GA 761727).